\documentclass[%
 preprint,
 superscriptaddress,
 amsmath,amssymb,
 aps,
 showkeys,
 titlepage,
 endfloats*   
]{revtex4-2}

\usepackage[utf8]{inputenc}
\usepackage[english]{babel}

\usepackage{graphicx}
\usepackage{dcolumn}
\usepackage{bm}
\usepackage[colorlinks = true,
            linkcolor = blue,
            urlcolor  = blue,
            citecolor = red,
            anchorcolor = blue]{hyperref}

\begin{document}


\title{\textit{in situ} Monitoring of Lithium Electrodeposition using Transient Grating Spectroscopy} 

\author{Runqing Yang}
\thanks{These authors contributed equally to this work.}
\affiliation{Department of Mechanical Engineering, University of California, Santa Barbara, CA 93106, USA}

\author{Harrison Szeto}
\thanks{These authors contributed equally to this work.}
\affiliation{Department of Chemistry and Biochemistry, University of California, Santa Barbara, CA 93106, USA}

\author{Brandon Zou}
\affiliation{Department of Mechanical Engineering, University of California, Santa Barbara, CA 93106, USA}

\author{Emily Spitaleri}
\affiliation{Department of Mechanical Engineering, University of California, Santa Barbara, CA 93106, USA}

\author{Bolin Liao}
\email{bliao@ucsb.edu} \affiliation{Department of Mechanical Engineering, University of California, Santa Barbara, CA 93106, USA}

\author{Yangying Zhu}
\email{yangying@ucsb.edu} \affiliation{Department of Mechanical Engineering, University of California, Santa Barbara, CA 93106, USA}

\date{\today}

\begin{abstract}
The mechanisms of lithium electrodeposition, which overwhelmingly affect lithium metal battery performance and safety, remain insufficiently understood due to its electrochemical complexity. Novel, non-destructive and \textit{in situ} techniques to probe electrochemical interfaces during lithium electrodeposition are highly desirable. In this work, we demonstrate the capability of transient grating spectroscopy to monitor lithium electrodeposition at the micrometer scale by generating and detecting surface acoustic waves that sensitively interact with the deposited lithium. Specifically, we show that the evolution of the frequency, velocity and damping rate of the surface acoustic waves strongly correlate with the lithium nucleation and growth process.
Our work illustrates the sensitivity of high-frequency surface acoustic waves to micrometer scale changes in electrochemical cells and establishes transient grating spectroscopy as a versatile platform for future \textit{in situ} investigation of electrochemical interfaces.
\end{abstract}

                            
\maketitle



Energy storage with rechargeable batteries is a critical enabler for renewable energy technologies and decarbonization~\cite{larcher2015towardsintro,goodenough2010challengesintro,liu2019pathways}. Despite extensive efforts to develop new materials~\cite{manthiram2020reflection} and nanostructures~\cite{chan2008high, boebinger2020spontaneous} to improve the energy density, power density, and fast-charging capability of rechargeable batteries~\cite{kang2009batterycederintro, liu2019challenges}, capacity decay ~\cite{severson2019data, wang2011grapheneyicuiintro} and thermal runaway ~\cite{goodenough2010challengesintro, feng2020mitigating} are still outstanding performance and safety challenges. For typical lithium-ion (Li-ion) and lithium-metal (Li-metal) batteries, degradation is usually accompanied with changes in the material properties and microstructures of the electrodes. Examples include lithium plating on graphite anodes~\cite{liu2016understandingplating}, the formation of cracks associated with large volume changes during cycling~\cite{bhattacharya2011transmission}, lithium dendrites penetrating the separator~\cite{li2015synergeticdendrites,li2018suppression}, and the formation of dead lithium~\cite{chen2017dead}. 
Due to the reactive nature of battery materials when exposed to air and moisture~\cite{boebinger2020understanding}, significant efforts have focused on developing \textit{in situ} and \textit{operando} experimental techniques, such as transmission electron microscopy (TEM)~\cite{huang2010situTEM,mcdowell2012effectTEM}, X-ray imaging and spectroscopy~\cite{eastwood2014lithiationXR,ebner2013visualizationXR} and nuclear magnetic resonance (NMR)~\cite{key2009realNMR,trease2011situNMR}, for microscopic characterization of battery materials. These developments have produced a wealth of insights into the fundamental reaction mechanisms of different electrode materials, but they can still be challenging because of ultrahigh vacuum requirements and/or time-consuming sample preparation and measurement.

As another, non-destructive, \textit{in situ} method with less stringent experimental conditions, ultrasonic acoustic characterization of battery materials has recently received increased attention. Hsieh \textit{et al.} demonstrated the capability of ultrasonic time-of-flight measurement to probe the physical dynamics of a Li-ion battery during cycling~\cite{hsieh2015electrochemical}. Similarly, Gold \textit{et al.}~\cite{gold2017probingpreSAW} used a lower-frequency
ultrasonic pulse (200\,kHz versus 2.25\,MHz in ~\cite{hsieh2015electrochemical}) to analyze the arrival time of the slow
compressional wave. These works demonstrate that the simple, \textit{operando} acoustic technique can be utilized to detect Li plating in commercial-scale Li-ion batteries as more significant Li plating results in increased hysteresis of the acoustic time of flight~\cite{davies2017statepreSAW,gold2017probingpreSAW,hsieh2015electrochemical}.
However, given the low frequency and the long wavelength of the acoustic waves used in these measurements, the spatial resolution is very limited and usually the entire battery is studied, while important phenomena such as initial Li nucleation and Li plating happen on microscopic length scales that cannot be resolved by the conventional ultrasonic techniques.
 Thus, an acoustic method with significantly increased spatial resolution is necessary to improve understanding of the microscopic, electrochemical processes occurring in Li-ion batteries. 
 
 In this work, we propose the use of transient grating spectroscopy (TGS)~\cite{choudhry2021characterizingreview}, which generates and detects high-frequency surface acoustic waves (SAWs) with tunable wavelengths in the micrometer range to probe the microscopic evolution of Li nucleation, growth and propagation on copper within a custom, \textit{in situ} cell. We choose to study Li deposition dynamics as understanding factors affecting its nucleation and growth are critical for the safety of the existing Li-ion batteries as well as potentially enabling commercialization of batteries with a high capacity, Li metal anode. TGS utilizes a pair of optical pump beams that cross with a controlled angle on the sample surface at the same location, where the interference of the two pump beams forms a one-dimensional (1D) periodic intensity distribution. Absorption of this optical excitation by the sample surface leads to a 1D periodic thermal stress that initiates a SAW with a wavelength matching the periodicity of the 1D excitation. The 1D periodicity of the optical excitation (the ``grating period''), and thus the SAW wavelength, can be conveniently tuned by the pump crossing angle and scanned from a few to tens of micrometers, which sets the spatial resolution of TGS. Subsequently, the propagation and decay of the SAW can be monitored by measuring the diffraction of a probe beam by the SAW, which encodes information about both the mechanical properties and the distribution of defects in the sample~\cite{PhysRevLett.111.036103resonance,reza2020non}. Here, we demonstrate that the SAW response in TGS is highly sensitive to the Li electrodeposition process in an \textit{in situ} cell. Specifically, we found that the evolution of the SAW frequency and damping strongly correlate with the stages of Li growth. Our result suggests that the TGS technique can provide a powerful platform for future \textit{in situ} investigation of Li nucleation and growth and potentially other surface/interface processes during battery operations with a high spatial resolution.  

Figure~1(a) shows a schematic of our TGS setup. More details of the setup can be found in the Supplementary Material. Briefly, a 515\,nm ns-pulsed pump beam is split by a phase mask with a switchable grating period and recombined at the sample with a controlled crossing angle through a telescope system to generate a SAW. A 532\,nm continuous-wave probe beam is diffracted by the SAW on the sample surface, and the diffracted intensity is monitored by a photodiode. The diffracted probe beam is coherently superposed with a reference beam, whose phase can be fine-tuned to realize heterodyne detection that increases sensitivity~\cite{choudhry2021characterizingreview}.  A schematic of the sample structure is shown in Fig.~1(b). The sample is a standard coin cell modified with an optical window (a 170\,$\mu$m glass) to allow TGS access. A 50-100 nm copper layer (Cu) was thermally evaporated on the backside of the glass, which serves as both the current collector for Li deposition and as the optical transducer for TGS owing to its optically smooth surface and high reflectivity. In our experiment, the TGS pump pulses are absorbed by the Cu layer and the resulting 1D periodic thermal stress generates a SAW that travels along the glass/Cu interface. The wavelength of the SAW matches the 1D periodicity of the pump excitation, which is controlled by the crossing angle of the two pump beams. The grating period on the sample is scanned between 10 to 20\,$\mu$m in this work. Although the SAW is bound to the glass/Cu interface, the associated elastic deformation (the ``evanescent tail'' of the SAW) extends into the cell, providing sensitivity to material changes in the cell. The SAW oscillation as a function of time after pump excitation is then monitored by measuring the intensity of the diffracted probe beam.


\begin{figure}[!htb]
\includegraphics[width=1\textwidth]{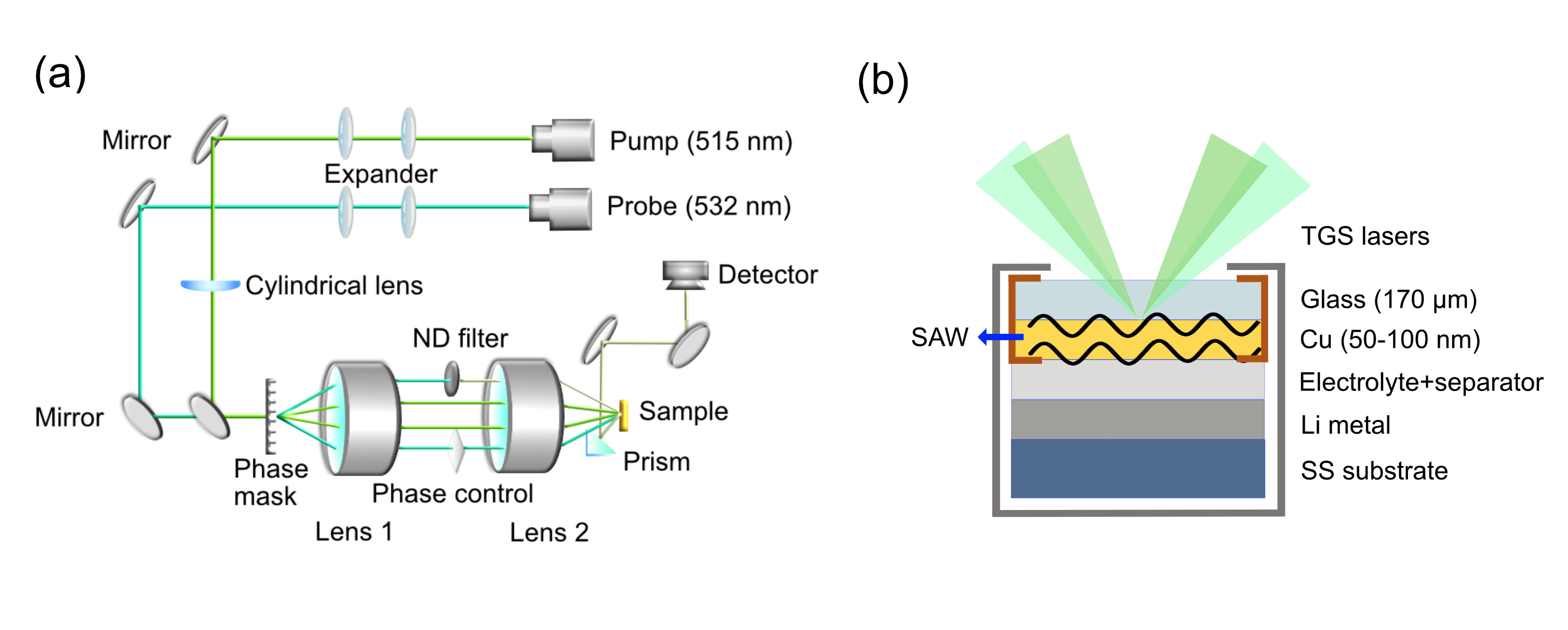}
\caption{\textbf{Schematic of the TGS setup and the sample cell structure. } (a) A schematic of the TGS setup used in this work. (b) A schematic of the structure of the sample optical cell studied in this work. Details of both the TGS setup and the sample cell structure can be found in the Supplementary Material. TGS: transient grating spectroscopy; SAW: surface acoustic wave; SS: stainless steel.} 
\label{fig:fig1}
\end{figure}


\begin{figure}[!htb]
\includegraphics[width=1\textwidth]{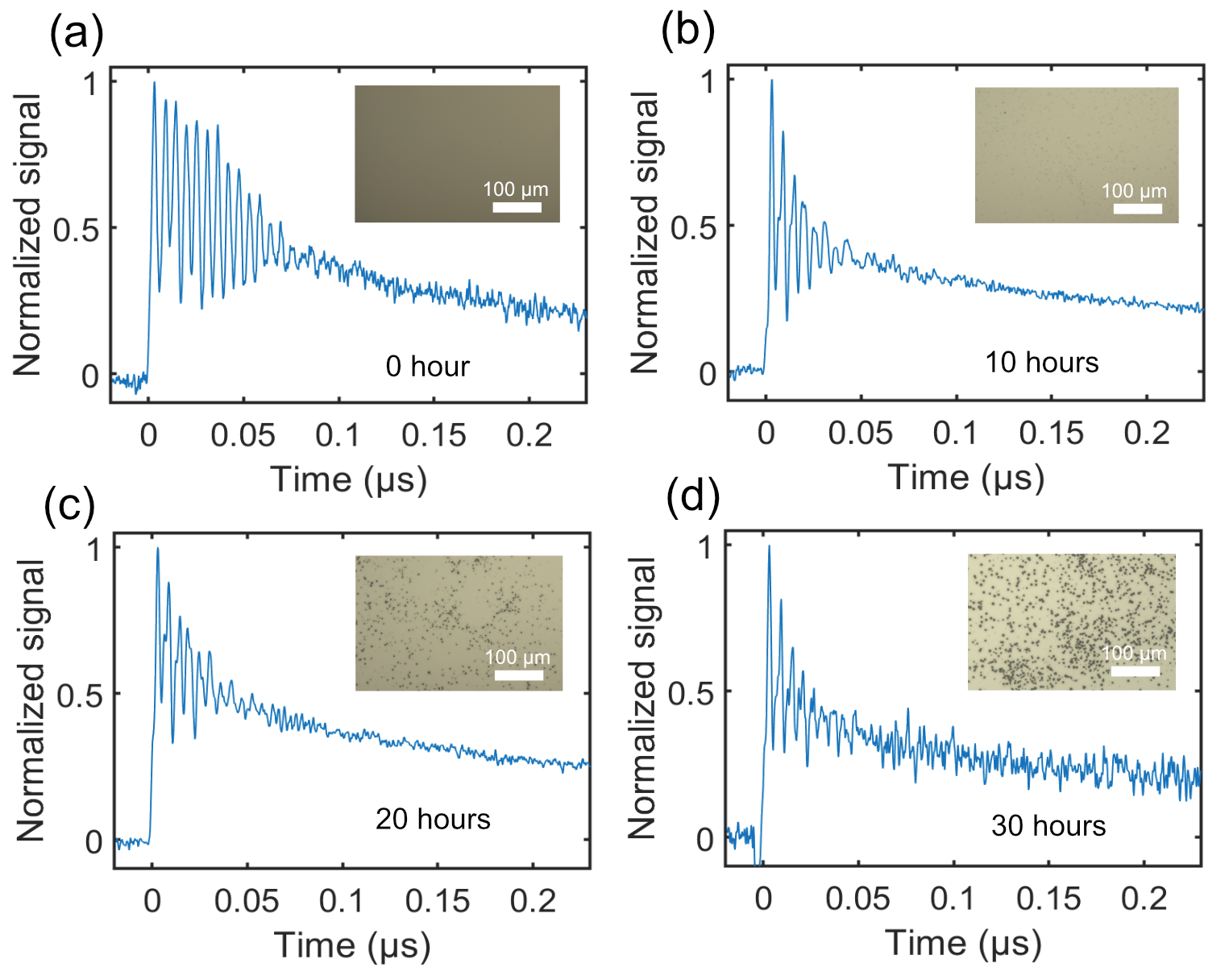}
\caption{\textbf{TGS traces and the corresponding optical microscope pictures during Li electrodeposition.} Normalized TGS signal taken with a grating period of 17.5\,$\mu$m on a sample cell with a 100-nm Cu layer (a) prior to Li electrodeposition, (b) after 10 hours of Li electrodeposition, (c) after 20 hours of Li electrodeposition and (d) after 30 hours of Li electrodeposition. The corresponding optical microscope pictures of the Cu layer are shown in the insets. A constant charging current of 88.42 $\mu$A/cm$^2$ was used for this dataset. } 
\label{fig:fig2}
\end{figure}

\begin{figure}[!htb]
\includegraphics[width=1\textwidth]{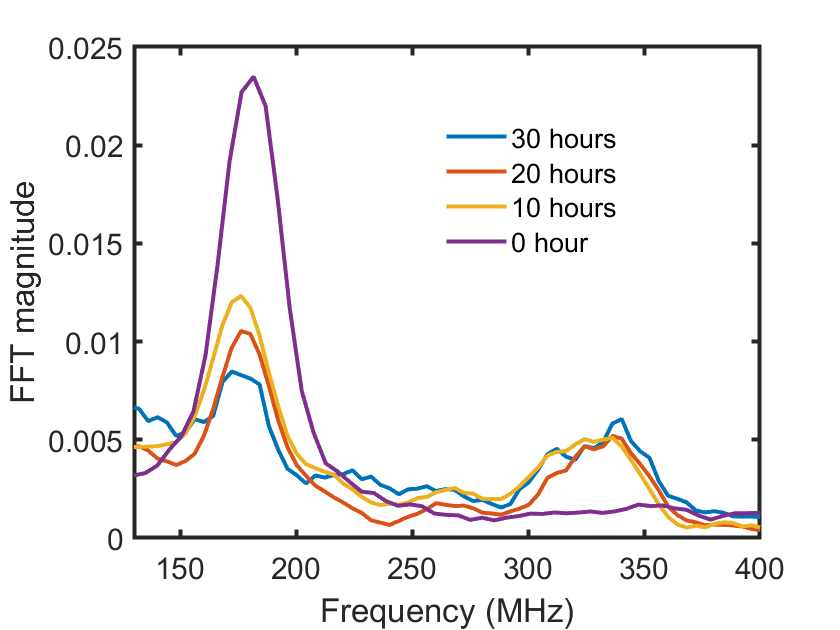}
\caption{\textbf{Fourier transform of the TGS traces.} Fast Fourier transform (FFT) magnitude of the raw TGS traces shown in Fig. 2 during the Li electrodeposition process. The low frequency peak at 180\,MHz weakens and broadens while a high frequency peak at 330\,MHz emerges as Li electrodeposition progresses. } 
\label{fig:fig3}
\end{figure}


To test the sensitivity of SAWs to material changes in the cell induced by electrochemical processes, we recorded the TGS signal from the sample at different stages of Li electrodeposition. The results taken with a grating period of 17.5\,$\mu$m are presented in Fig.~2, with the corresponding optical micrographs of the Cu surface shown in the insets. The thickness of the Cu layer used for this data set was 100\,nm. Before the electrodeposition process, the TGS signal is shown in Fig.~2(a). A clear and intense single-frequency SAW oscillation is observed, which is superposed on a decaying baseline signal that reflects the thermal diffusion process in the cell. Then we initiated the electrodeposition process by applying a charging current of 88.42 $\mu$A/cm$^2$ (more details in the SM). After 10 hours, as shown in Fig.~2(b), Li started to nucleate at isolated sites on Cu as seen from the optical micrograph. Correspondingly, a new, high frequency SAW emerged and the intensity of the original, low frequency SAW decayed faster, signaling a stronger damping of the SAW, likely by the growing Li clusters. After 20 hours [Fig.~2(c)], both the low and high frequency SAWs were more quickly damped. These results are consistent with the continued nucleation and growth of Li, as also reflected in the optical micrograph. This trend continued until 30 hours after electrodeposition when we stopped our experiment. At this time, Li clusters accumulated to form large islands and particles, which led to heterogeneity and roughness at the interfaces that strongly scattered the SAWs and also introduced noise in the TGS signal. A scanning electron microscope (SEM) image of the deposited Li is provided in the SM. The overall damping rate also increased after the electrodeposition process with all grating periods. Our results showed that the SAW signal measured by TGS is highly sensitive to the Li nucleation and growth process on the Cu surface. 

In Fig.~3, we quantify the evolution of the SAW signal during Li electrodeposition by conducting a Fourier transform of the TGS signal shown in Fig.~2. Prior to electrodeposition, a single SAW frequency of around 180\,MHz was observed. We note here that the SAW frequencies obtained in TGS are much higher than those used in previous ultrasonic measurements (typically MHz or below), indicating a much shorter wavelength and higher spatial resolution. After 10 hours of electrodeposition, a new, high frequency SAW peak emerged around 330\,MHz, consistent with the observed TGS signal in Fig.~2(b). Meanwhile, the width of the frequency peak at 180\,MHz, reflected by its full width at half-maximum (FWHM), broadened as a result of increased damping of the SAW. After 20 hours and 30 hours of electrodeposition, the intensity of the 180-MHz peak continued to decrease with a broadened peak width while the intensity of the 330-MHz peak remained relatively constant.

One advantage of TGS is the capability to readily change the periodicity of the optical pump excitation, and thus the SAW wavelength, by controlling the crossing angle of the pump beams. From the wavelength-dependent SAW frequency, the velocity of the SAW can be extracted. In principle, the spatial distribution of defects, which act as scatterers of the SAW, can also be mapped by examining the wavelength-dependent SAW damping. To test this capability, we varied the grating period from 13\,$\mu$m to 17.5\,$\mu$m when measuring a cell with a 50 nm copper layer. The raw TGS signal as a function of the grating period is provided in Section II of the Supplementary Material (SM). No qualitative difference was observed between the TGS signals of the 100-nm Cu sample and the 50-nm Cu sample. The data show that all of the TGS grating periods used can detect the first Li nucleation at almost the same time, since the Li film is thin relative to the gating period. Figure~4(a) shows the frequencies of the two SAW oscillations after 10 hours of Li electrodeposition as a function of the inverse grating period. Linear dependence for both SAW frequencies is observed, the slopes of which represent the velocity of the corresponding SAWs. The lower frequency corresponds to a velocity around 3000\,m/s, whereas the higher frequency corresponds to a velocity around 6000\,m/s. The lower-velocity SAW can be attributed to the Rayleigh-Lamb wave propagating along the surface of the Cu-coated glass, whose velocity is close to that of the Rayleigh wave in glass because of the small Cu thickness. The origin of the high-frequency SAW is more complicated. Based on the observation that this mode only emerges after Li electrodeposition, we tend to attribute it to an acoustic mode in Li since its velocity is very close to the literature value of the acoustic velocity in Li (around 6000\,m/s) ~\cite{Lisoundspeed}. An alternative explanation for this mode is the so-called ``surface-skimming'' longitudinal acoustic mode in glass with a similar velocity that is sometimes observed in TGS~\cite{maznev1997time}. However, it is unclear how the intensity of this mode detected in TGS can be enhanced by the presence of Li clusters on Cu surface. The exact origin of this high-frequency SAW will be investigated in our future work. Regardless of its origin, the emergence of this new SAW frequency can serve as a sensitive probe for the beginning of Li nucleation and growth, which concurs with increased damping of the lower-frequency SAW.

To quantify the increased damping of the lower-frequency SAW, we can analyze the damping coefficient $\Gamma_A$ of this mode as a function of the electrodeposition time and the grating period. In order to ensure that the intrinsic acoustic damping is measured in the TGS and avoid the geometric damping effect due to a finite optical beam size, we used a cylindrical lens to focus the optical pump beams to obtain an illuminated area on the sample that is long along the SAW propagation direction~\cite{amirkhani2011polymer} (more details in the Supporting Material). $\Gamma_A$ can be extracted by fitting the experimental TGS traces using the following equation~\cite{cucini2010acousticequation}:
\begin{align}
S(t)=&Ae^{-\Gamma_A t}\cos(\omega_At)+Be^{-\Gamma_At}\sin(\omega_At)+Ce^{-\Gamma_H t},
\end{align}
where $S(t)$ is the TGS trace and the SAW angular frequency $\omega_A$ can be determined by the Fourier Transform analysis. The remaining fitting parameters are: the SAW damping coefficient $\Gamma_A$, the thermal decay rate $\Gamma_H$, and the corresponding amplitude constants A, B and C. Here, we focus on the damping coefficient of the lower-frequency SAW. The damping rate $\Gamma_A$ as a function of the electrodeposition time and the grating period is shown in Fig.~4(b). Our results show that $\Gamma_A$ increases with the electrodeposition time as more Li nucleates and grows on the Cu surface. The Li clusters serve as scattering centers for the acoustic wave and an increased density of Li clusters contributes to stronger acoustic damping. In the meantime, no clear dependence of the acoustic damping coefficient on the grating period was observed, suggesting that the spatial distribution of Li clusters was relatively uniform on the length scale of the grating periods used.   


\begin{figure}[!htb]
\includegraphics[width=1.1\textwidth]{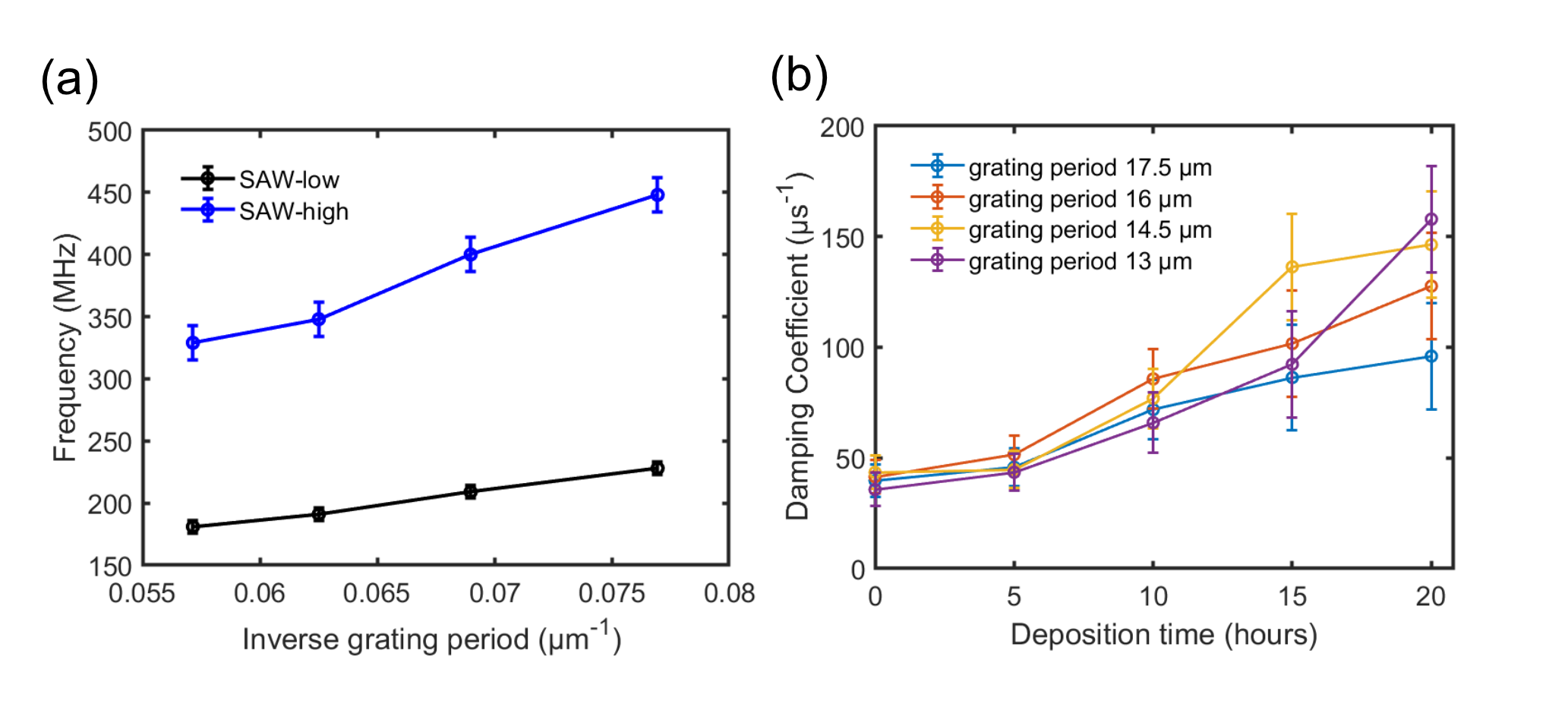}
\caption{\textbf{SAW frequency and damping rate as a function of TGS grating period.} (a)The relation between the inverse grating period used for the TGS measurement and the measured frequencies of the two SAW oscillations. The slope represents the velocity of the corresponding SAW mode. (b)The damping coefficient of the SAW at 180\,MHz as a function of the Li electrodeposition time and the grating period used in TGS.} 
\label{fig:fig4}
\end{figure}

In summary, our study demonstrates that the SAW generated and detected in TGS can be used to investigate the electrodeposition behaviors of Li in an \textit{in situ} electrochemical cell. Our experimental results, along with optical microscope images, indicate that TGS is capable of detecting micrometer-level changes on the Cu surface as a result of Li electrodeposition. The frequencies and damping rates of the SAW in the TGS signal can be used to sensitively monitor the initial nucleation and growth of Li. Our work establishes TGS as a versatile platform for \textit{in situ} investigations of electrochemical processes as well as potential studies of surface/interface cracks, defects, and dendrites in Li ion batteries.

Additional details of the experimental methods and additional data are available in the Supplementary Material.

We thank Dr.\,Alexei Maznev for helpful discussions. This work was supported by an Early Career Faculty grant from NASA’s Space Technology Research Grants Program under award number 80NSSC23K0072. We thank Thomas Miller and Richard Oeftering at NASA Glenn Research Center for valuable feedback on this work. The TGS setup development was supported by a DURIP grant from the U.S. Army Research Office under the award number W911NF-20-1-0161. R. Yang also acknowledges the support of a Chancellor's Fellowship at the University of California, Santa Barbara. A portion of this work was performed in the UCSB Nanofabrication Facility, an open-access laboratory.

\section*{Conflict of Interest Statement}
The authors have no conflicts to disclose.

\section*{Author Contributions}
Y.Z. and B.L. conceived and supervised the project. H.S., B.Z. and E.S. fabricated the cells and conducted electrochemical characterizations. R.Y. conducted the TGS measurement. R.Y.,B.L. and Y.Z. drafted the manuscript. All authors read and edited the manuscript.

\section*{Data Availability Statement}
The data that support the findings of this study are available within the article and its supplementary material. Additional data are available from the corresponding authors upon reasonable request.
\bibliography{references.bib}

\end{document}



\title{Supplementary Material: In Situ Monitoring of Lithium Electrodeposition using Transient Grating Spectroscopy} 

\author{Runqing Yang}
\thanks{These authors contributed equally to this work.}
\affiliation{Department of Mechanical Engineering, University of California, Santa Barbara, CA 93106, USA}

\author{Harrison Szeto}
\thanks{These authors contributed equally to this work.}
\affiliation{Department of Chemistry and Biochemistry, University of California, Santa Barbara, CA 93106, USA}

\author{Brandon Zou}
\affiliation{Department of Mechanical Engineering, University of California, Santa Barbara, CA 93106, USA}

\author{Emily Spitaleri}
\affiliation{Department of Mechanical Engineering, University of California, Santa Barbara, CA 93106, USA}

\author{Bolin Liao}
\email{bliao@ucsb.edu} \affiliation{Department of Mechanical Engineering, University of California, Santa Barbara, CA 93106, USA}

\author{Yangying Zhu}
\email{yangying@ucsb.edu} \affiliation{Department of Mechanical Engineering, University of California, Santa Barbara, CA 93106, USA}

\date{\today}

\maketitle



\section{Experimental Methods}

\subsection{\textit{in situ} TGS cell fabrication}
In order to probe Li electrodeposition with TGS, a glass coverslip (Thorlabs $\varnothing$12 mm  $\#$1.5H 170 $\pm$ 5 µm, Item $\#$ CG15NH1)  was first cleaned successively in acetone, isopropanol and deionized water for 15 minutes each before being dried with dry N$_2$ and placed in an O$_2$ plasma cleaner for 3 minutes. The glass coverslip was subsequently brought into a thermal evaporator where a 5-nm Cr adhesion layer and 50/100-nm Cu thin film were deposited at a pressure of $3 \times 10^{-6}$ Torr and rates of 1 and 5\,\AA/s, respectively. The glass coverslip was subsequently epoxied (Loctite EA E-120HP) to the top cap of a SS304 20232 coin cell case (MTI Corporation, Item $\#$ CR2032CASE304), which had a 7/32'' hole punched out to provide optical access. After allowing the epoxy to cure for a day, Cu foil tabs were spot welded to the outer edge of the coin cell top case in order to establish electrical contact between the Cu thin film and coin cell case. The case was subsequently transferred into an Ar glovebox with H$_2$O and O$_2$ levels below 1\,ppm to fabricate an \textit{in situ} TG cell. 

\subsection{Cell Assembly}
In order to accommodate the Cu-coated glass coverslip, an \textit{in situ} coin cell was prepared in an inverted, half cell architecture. As shown in Fig.~1(b) in the main text, the \textit{in situ} cell consists of a Cu-coated glass coverslip, Celgard 2325 separator, Li counter/reference electrode (MTI Corporation), SS304 spacer (MTI Corporation), and SS304 wave spring (MTI Corporation) along with 300\,$\mu$L 1\,M LiPF$_6$ in 1:1 by vol. EC:DMC (Sigma Aldrich) electrolyte.

\subsection{Li electrodeposition}
A Biologic SP-150 potentiostat was used to apply a galvanostatic current of 50 and 100 $\mu$A in order to deposit Li on the Cu surface. The current density was estimated to be 44.21 and 88.42 $\mu$A$cm^{-2}$ for 50 and 100-nm Cu samples based on the 12-mm diameter of the glass coverslip. The \textit{in situ} cell was periodically taken off of the potentiostat in order to take \textit{in situ} TGS and optical microscopy measurements. Potential of Cu electrode vs. a Li reference electrode during the first 5 hours of electrodeposition is shown in Fig.~\ref{fig:potential}. The cell potential quickly decreases and remains below 0 V vs. $\textrm{Li}^{+}$/Li which indicates Li electrodeposition is occurring.

\subsection{Transient Grating Spectroscopy}
In the TGS setup, illustrated schematically in Fig.~1(a), an optical diffraction grating (``phase mask'') separates the pump and probe beams into $\pm$1 diffraction orders, and a two-lens confocal imaging system is employed to recombine the pulses at the sample at a controlled crossing angle $\theta$. The two crossed pump beams with the crossing angle $\theta$ creates an interference pattern with a period of $L = \lambda/[2\sin(\theta/2)]$, where $\lambda$ is the wavelength of the pump light. The decay in the intensity of the transient gratings is monitored by diffraction of a probe laser beam, and the grating period will determine wavelength of the generated SAW, which, in our study, was varied between 13 and 17.5\,$\mu$m. The TGS employs optical heterodyne detection.  One of the 1st order probe beams is attenuated with a neutral density filter to serve as a reference beam. The diffracted signal is coherently superposed with the reference beam, and the reference beam is directed into a fast silicon avalanche photodiode (Hamamatsu C5658). The relative phase of the probe beam and the reference beam can be controlled by a separate phase plate (in this work we used a transparent plate with highly parallel surfaces) in the probe path. The signal is subsequently read out using an oscilloscope (Tektronix TDS784A). Heterodyne detection boosts the signal level as well as yields a signal that is linear with respect to the material response, simplifying the analysis and interpretation of the measurements. The pump has a 515-nm wavelength, a 1.3-ns pulse width, and a tunable repetition rate up to 2\,kHz (Coherent Flare NX). A continuous-wave 532-nm laser (Coherent Sapphire SF NX) is used as the probe beam. The time resolution of this TGS implementation is limited by the bandwidths of the photodetector and the oscilloscope and estimated to be a few nanoseconds. The pump pulses of 515-nm wavelength, 1.3$\pm0.2$ ns pulse
width and a
repetition rate of 2 kHz (Coherent Flare NX 515-0.6-2) have a fluence of 10 $\mu$J/cm$^2$ and are focused by a cylindrical lens to a line profile (a length of $\sim$5\,mm along the grating direction and a width of $\sim$220\,$\mu$m). This line profile is to avoid geometric damping of the SAW by the finite beam size along the grating direction. The continuous-wave (CW) 532-nm probe beam has a power of 14\,mW/cm$^2$ with a diameter of 200 $\mu$m.

\subsection{Optical microscopy}
A Nikon Eclipse Upright Microscope was used to examine changes at the Cu interface (through the glass coverslip and Cu thin film).

\subsection{Scanning Electron Microscopy}
After TGS and optical microscopy measurements, the cell was transferred back into an Ar glovebox for cell disassembly. Upon disassembly, the electrodeposited Li on Cu/glass substrate was washed three time with DMC in order to remove residual electrolyte and left to dry overnight. The dried Li on Cu/glass substrate was sealed inside the Ar glovebox and subsequently transferred to an FEI XL-30 SEM with minimal exposure to the environment. An SEM image of the deposited Li after 30 hours of electrodeposition is shown in Fig.~\ref{fig:SEM}.

\clearpage

\section{Additional Data}
\begin{figure}[!htb]
\includegraphics[width=0.75\textwidth]{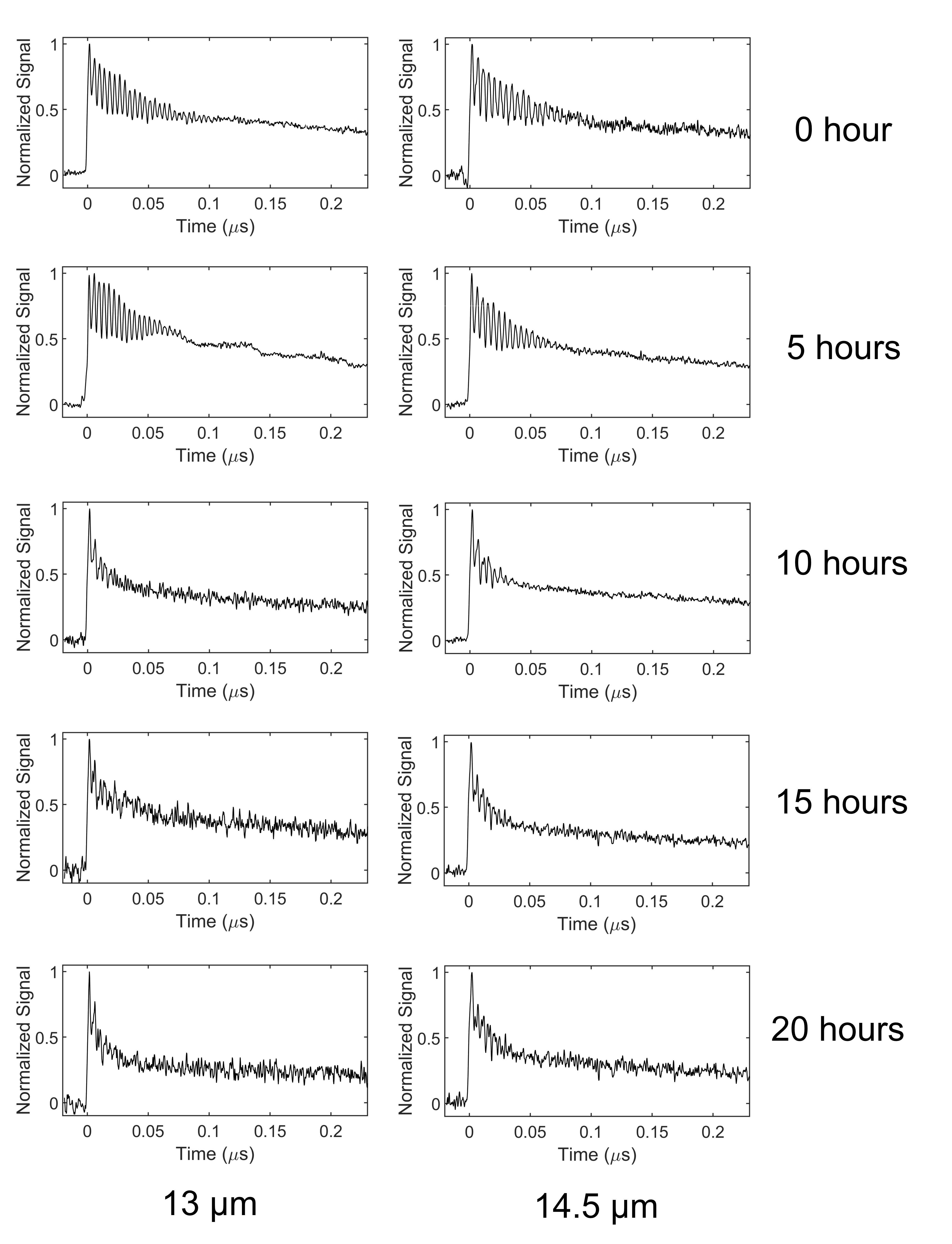}
\caption{\textbf{TGS traces taken with grating periods of 13\,$\mu$m and 14.5\,$\mu$m on a sample cell with a 50-nm Cu layer}. The TGS traces were taken after 0, 5, 10, 15 and 20 hours of Li electrodeposition.} 
\label{fig:appendix1}
\end{figure}
\clearpage

\begin{figure}[!htb]
\includegraphics[width=0.75\textwidth]{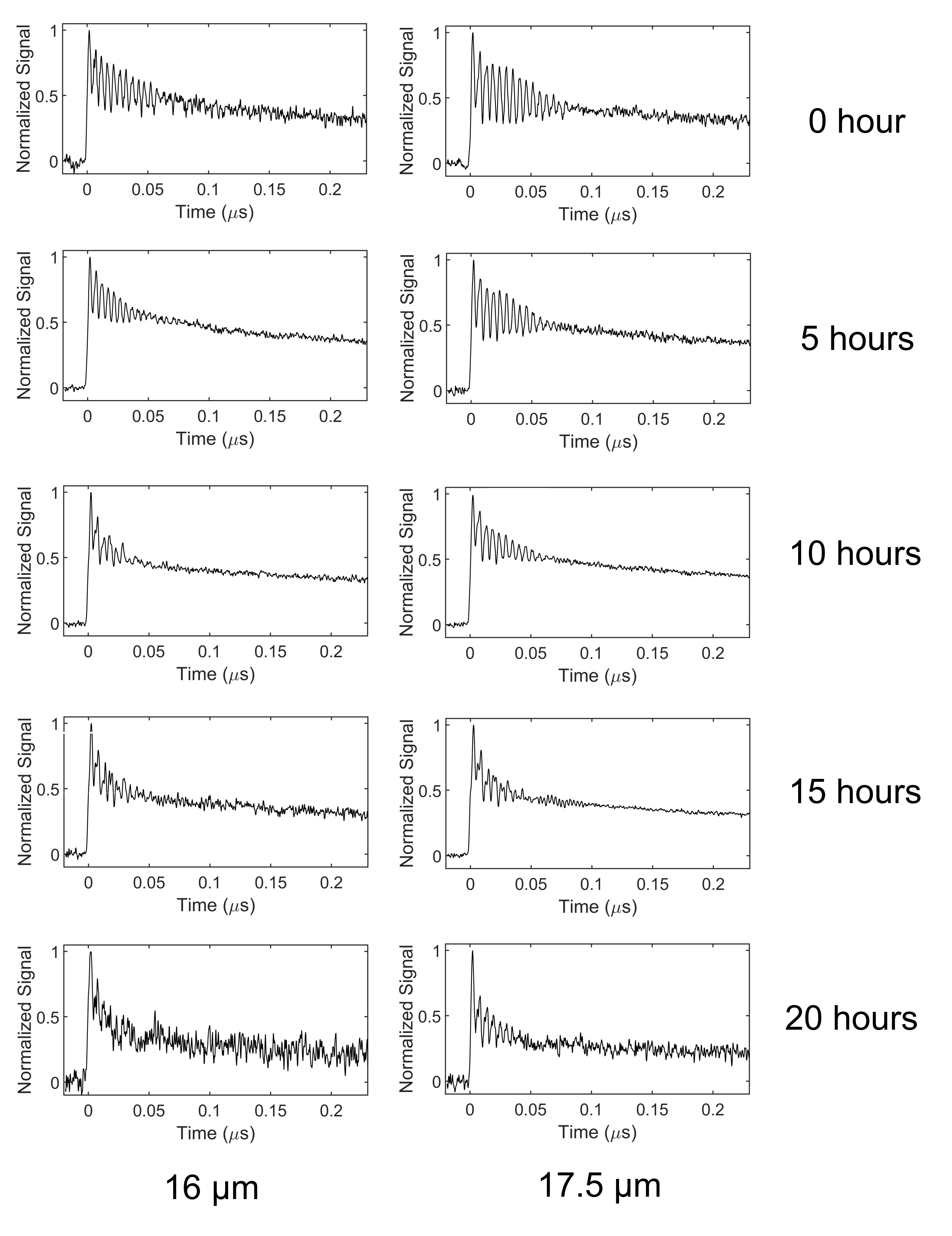}
\caption{\textbf{TGS traces taken with grating periods of 16\,$\mu$m and 17.5\,$\mu$m on a sample cell with a 50-nm Cu layer}. The TGS traces were taken after 0, 5, 10, 15 and 20 hours of Li electrodeposition.} 
\label{fig:appendix2}
\end{figure}
\clearpage

\begin{figure}[!htb]
\includegraphics[width=1\textwidth]{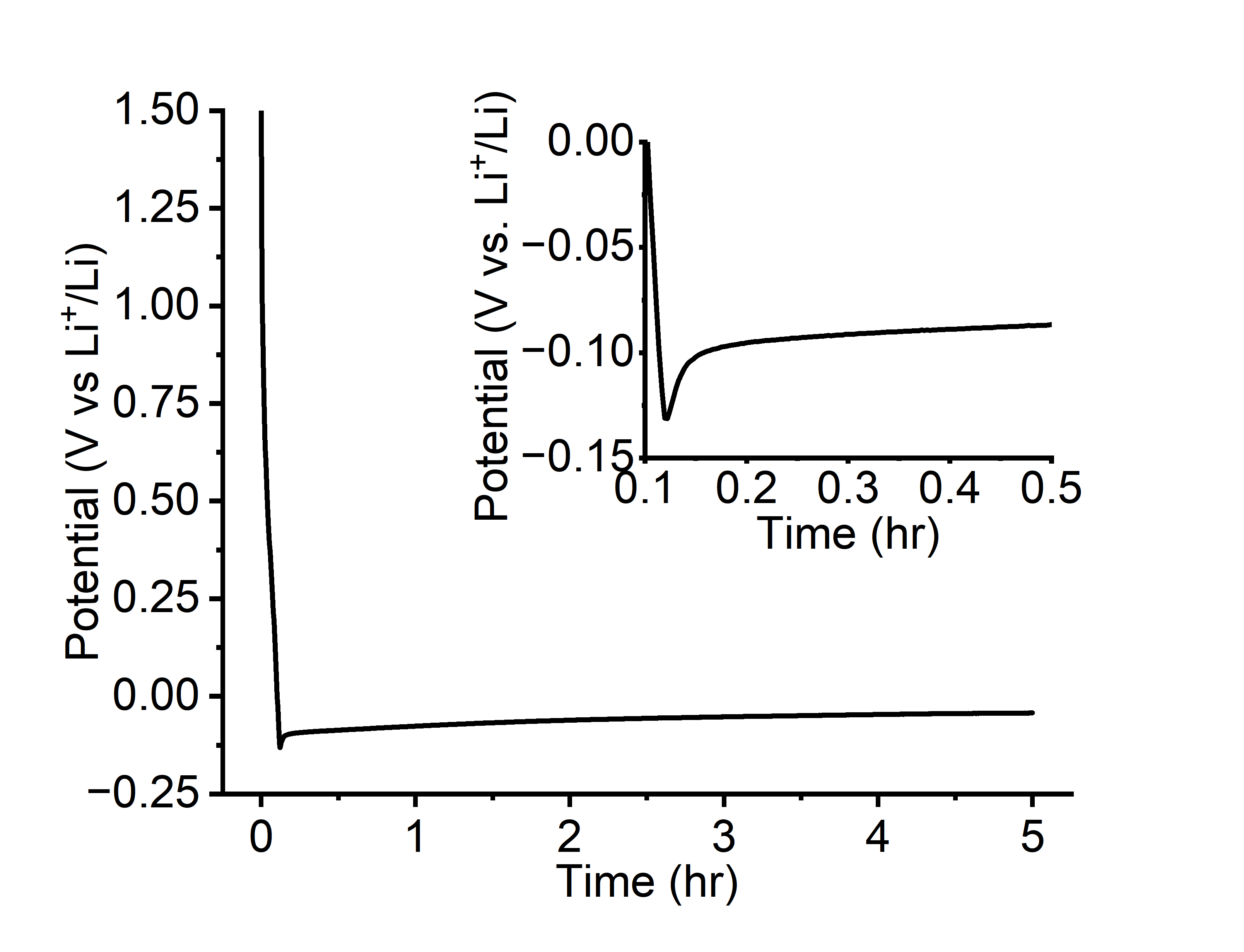}
\caption{\textbf{Potential of the Cu electrode vs. a Li reference electrode during first 5 hours of electrodeposition.}} 
\label{fig:potential}
\end{figure}
\clearpage

\begin{figure}[!htb]
\includegraphics[width=1\textwidth]{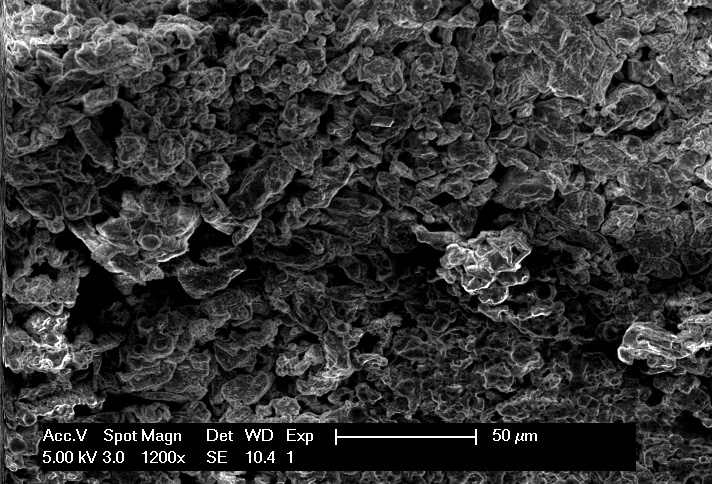}
\caption{\textbf{SEM image of the deposited Li on Cu surface after 30 hours of electrodeposition}. The applied current was 88.42 $\mu$A/cm$^2$.} 
\label{fig:SEM}
\end{figure}
